\documentclass[twocolumn]{aastex631}

\shorttitle{CR ionization rate in  PPD with sheared magnetic fields}
\shortauthors{Fujii \& Kimura}
\graphicspath{{./}{figures/}}

\begin{document}

\title{Cosmic-ray ionization rate in protoplanetary disks with sheared magnetic fields}

\author[0000-0002-3648-0507]{Yuri I. Fujii}
\affiliation{Graduate School of Human and Environmental Studies,\\
    Kyoto University, Yoshida-Nihonmatsu, Sakyo, Kyoto 606-8501, Japan}
    \affiliation{Department of Physics, Nagoya University, Furo-cho, Chikusa-ku, Nagoya, Aichi 464-8602}

\author[0000-0003-2579-7266]{Shigeo S. Kimura}
\affiliation{Frontier Research Institute for Interdisciplinary Sciences,\\
 Tohoku University, Sendai 980-8578, Japan}
\affiliation{Astronomical Institute, Graduate School of Science,\\ Tohoku University, Sendai 980-8578, Japan}

\begin{abstract}
We investigate the effects of magnetic field configurations on the ionization rate by cosmic rays in protoplanetary disks. First, we consider cosmic-ray propagation from the interstellar medium (ISM) to the protoplanetary disks and showed that the cosmic-ray density around the disk should be 2 times lower than the ISM value. Then, we compute the attenuation of cosmic rays in protoplanetary disks. The magnetic fields in the disk are stretched to the azimuthal directions, and cosmic rays need to detour while propagating to the midplane. %We 
Our results show that the detouring effectively enhances the column density by about two orders of magnitudes. 
We employ a typical ionization rate by cosmic rays in diffuse ISM, which is considered too high to be consistent with observations of protoplanetary disks, and find that
the cosmic rays are significantly shielded at the midplane.
In the case of the disk around IM lup, the midplane ionization rate is very low for $r\lesssim\,100$ au, while the value is as large as a diffuse ISM in the outer radii.
Our results are consistent with the recent ALMA observation that indicates the radial gradient in the cosmic-ray ionization rate. The high ionization rate in the outer radii of disks may activate the magnetorotational instability that was thought to be suppressed due to ambipolar diffusion. 
These results will have a strong influence on the dynamical and chemical evolutions of protoplanetary disks.
\end{abstract}


\section{Introduction} \label{sec:intro}
Protoplanetary disks are the birthplaces of planets. They consist of weakly ionized gas, and their dynamics are mainly governed by the magnetohydrodynamic (MHD) processes. Hence, the ionization degree inside the disks determines the fate of the protoplanetary disks \citep[e.g.,][]{san00, war07, gle12, gre15, bet17, bai17, mor19}. Ionization also affects the chemical reactions in the disk, which is crucial to interpreting the observational data of ALMA \citep[e.g.,][]{wal12, eis16, Aikawa21maps, law21, obe21, not21}

\begin{figure}
   \plotone{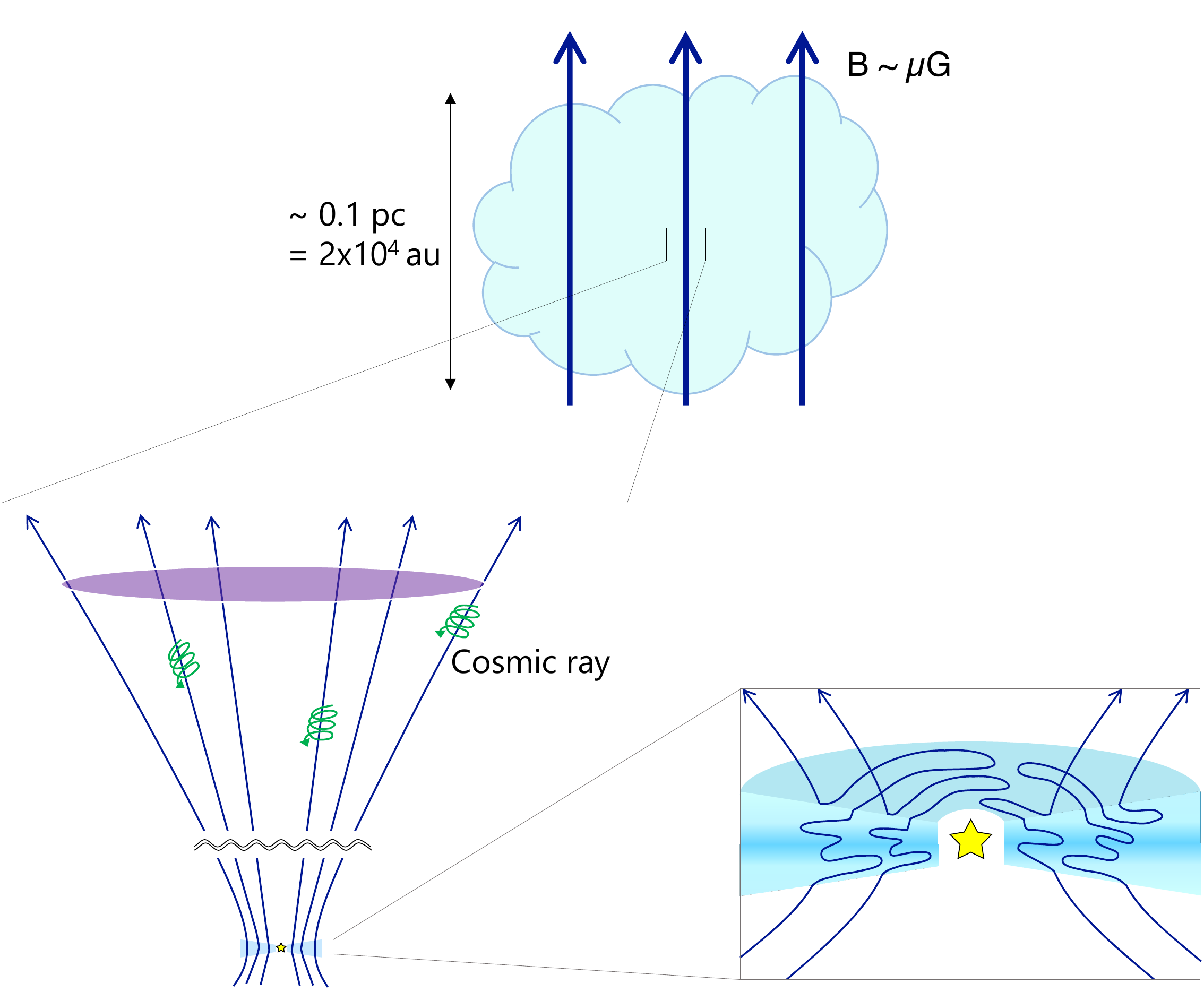}
   \caption{Schematic view of our scenario. An hourglass-shaped magnetic field configuration focuses cosmic rays from a molecular cloud core to a protoplanetary disk. The purple oval represents the area from which the disk can receive cosmic ray particles. The magnetic fields are stretched to azimuthal direction by shearing motion.
           }
               \label{fig:core_ppd}
    \end{figure}

Cosmic rays are an essential ionization source in protoplanetary disks. Protostellar radiations, including ultraviolet and X-rays, are the most powerful ionization source in the upper layer of the disk. However, they are easily blocked by the disk gas, and their ionization rates at the midplane are strongly suppressed \citep{ige99, erc13}. Cosmic rays have a stronger penetration power, and thus, they are expected to be the dominant source of ionization at the midplane of the inner disk \citep[e.g.,][]{gam96a, san00, war07, oku09, fuj11}, although many previous studies on ionization degree of protoplanetary disks only consider X-ray ionization \citep[e.g.,][]{gla97, ilg06a, fro02, tur08}.
In the deepest part of the disk, where even cosmic rays cannot penetrate, the ionization floor is determined by radionuclides.
The ionization rate by cosmic rays in protoplanetary disks is still controversial due to the uncertainty of cosmic-ray intensity there. \citet{cle13a, cle14a, cle15} discussed the modulation of cosmic rays by stellar winds from T-Tauri stars, i.e., T-Tauriosphere.
Cosmic rays can be produced by protostellar activities, which are known as stellar energetic particles. These may also be an important ionization source, especially in the surface layer of the disk \citep{tur09, pad15, pad16, rab17,fra18}.

Most of the previous studies assumed that the energy density of cosmic rays at the surface of a protoplanetary disk is the same as that in an interstellar medium \citep[e.g.,][]{gam96a, san00, fur13}.
They also consider that cosmic rays travel to the disk midplane in the shortest path, i.e., cosmic rays enter the protoplanetary disk straightly from the vertical direction. 
In reality, the shear motion in protoplanetary disks stretches the magnetic field, and cosmic rays travel only along the magnetic field lines due to their small gyration radii (see Figure~\ref{fig:core_ppd}).  Thus, the cosmic rays should detour to reach into the protoplanetary disks, which changes both the cosmic-ray energy density and the penetration depth. 
\citet{pad18} estimated the cosmic-ray ionization rate in the inner radii of a protoplanetary disk with taking the effect of the disk magnetic field into account.
In this Letter, we consider the magnetic fields that are connected to the molecular cloud and discuss the effects of  sheared field lines on the cosmic-ray ionization rate in global protoplanetary disks.

\section{Focusing and reflecting cosmic ray particles} \label{sec:2}

We consider the magnetic field configuration shown in Figure~\ref{fig:core_ppd}, where magnetic fields threading the protoplanetary disk are connected to its parent molecular cloud. This configuration is naturally expected if we consider disk formation by the gravitational collapse of the cloud core. Observations of some molecular cloud cores show an hourglass-shaped magnetic field configuration, which may support this picture~\citep[e.g.,][]{mau18}. This configuration is similar to those used by \citet{pad11, pad13}. In contrast, the magnetic fields in the solar wind extend horizontally with a spiral shape and are disconnected from those in the ISM~\citep{par58, pet19}.
If the magnetic field configuration around a protoplanetary disk is similar to that of the solar wind, 
the cosmic-ray density can be modulated due to T-Tauriosphere as discussed in \citet{cle13a, cle14a, cle15}. 
However, in the case that magnetic fields are connected to the ISM, cosmic rays can travel along the field lines
(see Section \ref{sec:discussion}). 

Gyro radius of a cosmic ray proton is  
$r_{\rm G}=E/(eB)\sim0.2~(E/10{\rm~GeV})(B/10{\rm~\mu G})^{-1}$ au, where $E$ is the energy of cosmic ray protons, $e$ is the elementary charge, $B$ is the magnetic field strength, and we use $B=10\rm~\mu G$ in a molecular cloud \citep[e.g.][]{cru10}. This is much smaller than the scale of the cloud 
core ($\sim 0.1$ pc $\sim 2\times10^4$ au), which can be regarded as the turbulence scale.\footnote{We consider that the kinetic-scale turbulence can be damped due to the low ionization rate in the cloud core and the disk. Also, the turbulence should be injected in the largest scale and the power spectrum of MHD turbulence is expected to be steep \citep{gol95}.} Thus, cosmic rays of $E\lesssim10$ TeV travel along magnetic field lines. 
We assume that cosmic-ray propagation is in the free-streaming regime and ignore the diffusion (see Section 5).
The disk bundles up the magnetic field lines from a large area\footnote{During the star and disk formation, a part of the magnetic fields slips out from the disk. We use the magnetic field strength after the disk formation in our estimate.}. If we assume that the cosmic-ray flux is proportional to the magnetic field strength, which is true for an isotropic cosmic-ray distribution and inefficient CR diffusion perpendicular to the magnetic field lines, the cosmic-ray flux at the disk surface is enhanced by a factor of 
\begin{equation}
    \frac{S_{\rm cloud}}{S_{\rm disk}} = \frac{B_{\rm disk}}{B_{\rm cloud}}\sim 10^3,
    \label{eq1}
\end{equation}
where $S_{\rm cloud}$ is the surface area in the molecular cloud from which the protoplanetary
disk can collect cosmic rays (the purple area in Figure~\ref{fig:core_ppd}), 
$S_{\rm disk}$ is the total area of the disk, and, 
$B_{\rm disk}\sim 10$ mG and $B_{\rm cloud}\sim 10\mu$G are the typical strengths of 
magnetic fields in a protoplanetary disk \citep[at $r\sim10$ au, $z\sim4H$, where $H$ is the scale height of the disk; cf.][]{suz10}) and molecular cloud core, respectively.

In accordance with the conservation of kinetic energy and magnetic moment of a cosmic ray particle,
pitch angles of the particles in a molecular cloud core, $\alpha_{\rm cloud}$, and 
that in a protoplanetary disk, $\alpha_{\rm disk}$, have the following relation \citep[e.g.][]{pad11}, 
\begin{equation}
    \frac{{\rm sin}^2\alpha_{\rm disk}}{{\rm sin}^2\alpha_{\rm cloud}}=\frac{B_{\rm disk}}{B_{\rm cloud}}.
\end{equation}
If we adopt the value in Equation (\ref{eq1}), 
the pitch angle in the cloud core should satisfy sin$\alpha_{\rm cloud}\leq0.03$, 
which means that only the cosmic ray particles that travel almost perfectly parallel to the field line 
can enter the protoplanetary disk. All the other particles are reflected by the magnetic mirror effect,
and the cosmic-ray flux is reduced by a factor of $\alpha^2/2$. Because of the balance between the focusing and reflecting effects, the net cosmic-ray flux at the disk surface ends up with a half of that in the parental cloud core \citep[see][for more detailed treatment]{sil18}.  This factor 2 is generic as long as $B_{\rm cloud} / B_{\rm disk} \ll 1$.
We define this factor for diminishing CR flux as $\Gamma_{\rm dim}$.

%%%%%%%%%%%%%%%%%%%%%%%%%%%%%%%%
\section{Disk Structure}
               \label{sec:disk}
%%%%%%%%%%%%%%%%%%%%%%%%%%%%%%%%
Below the disk surface, the magnetic fields are stretched to the azimuthal direction due to the shaer motion of the disk (see Figure~\ref{fig:core_ppd}).
In order to obtain the sheared field structure and the stratified density distribution of a protoplanetary disk, we perform magnetohydrodynamic
(MHD) simulations. 
The circumstance under which magnetorotational instability (MRI) generates turbulence is still controversial \citep{bai13b, bai13c, gre15, sim15b, mor19}. The activation of MRI is strongly affected by the ionization-rate distribution, which will be obtained in this work. In this Letter, we consider a non-turbulent protoplanetary disk because it leads to the most efficient ionization by cosmic rays (see Section \ref{sec:discussion} for the effect of turbulence).
We conduct low-resolution shearing-box 
simulations with ideal MHD approximation \citep{gol65, haw95} to mimic the non-turbulent disk structure.
%%%%%%%
\begin{figure}
%\epsscale{0.6}
   \centering
   \plotone{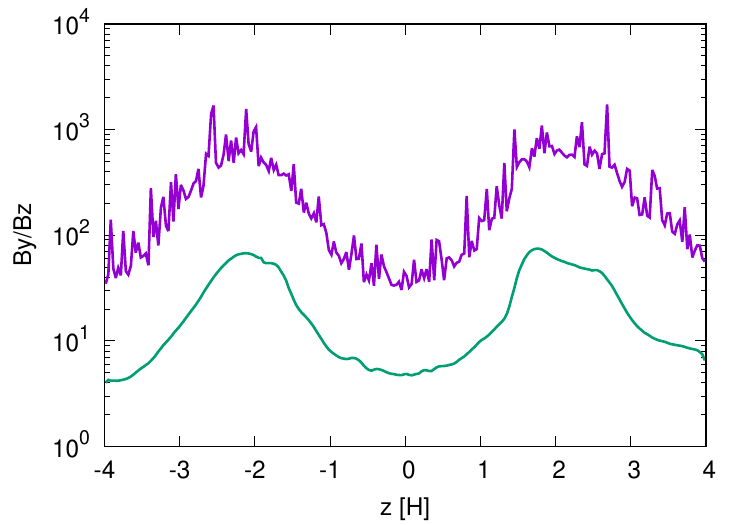}
   \plotone{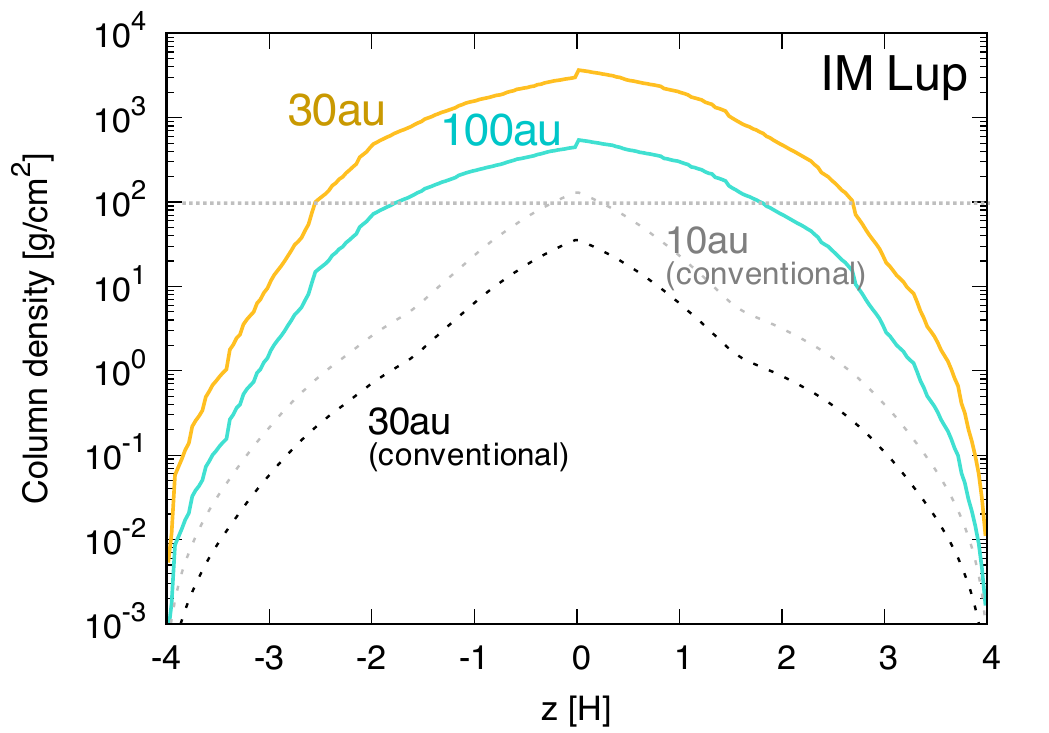}
   \caption{
       The top panel shows the ratio of the azimuthal and vertical components of the magnetic field
       strength at each height from the midplane. The purple jagged line represents 
       $\langle B_y/B_z\rangle$. We also plot $\langle B_y\rangle/\langle B_z\rangle$
       in green for comparison.
       The bottom panel presents the effective column densities that cosmic ray particles have to go through
       from the closer surface to reach the corresponding heights.
       The dashed curves are the vertical column densities that are used in the conventional estimates, and the dotted horizontal line represents the attenuation length of cosmic rays.
   }
               \label{fig:By_Bz}
    \end{figure}
%%%%%%%
The simulations were performed with Athena MHD code \citep{sto08, sto10}.
The benefit of the ideal MHD approximation is that simulations can be done with normalized physical values (Keplerian frequency, scale height of the disk, and the midplane density are set to be unity). We can freely scale the unit to arbitrary radii in the disk.
The resolution was $30\times60\times240$ for a local box with the size of 
$\pm 0.5H$, $\pm1H$, and $\pm4H$ in the $x$, $y$, and $z$ directions, respectively. 
The $z$-component of the initial plasma beta in the midplane of the disk 
(corresponding to the net vertical field) is parameterized as $\beta_{\rm z}=10^4$ and $ 10^6$.

The top panel of Figure~\ref{fig:By_Bz} shows the ratio of the $y$ and $z$ components of the magnetic fields as a function of the distance from the midplane for $\beta_{\rm z}=10^6$.
One can see that the shear elongates the field lines 
to the azimuthal direction by a factor of $100-1000$. This means that cosmic ray particles traveling along 
the magnetic fields have to go through  $\Gamma_{\rm detour}=\int\langle B_y/B_z\rangle dz\sim100-1000$ times longer distance 
to arrive at a certain height than the vertically straight path. Although the absolute values of the field strength are different between the cases with $\beta_{\rm z}=10^4$ and $10^6$, the values of $\Gamma_{\rm detour}$ are similar. Thus, our results can be applied to the disks that have typical magnetic fields ($\beta_{\rm z}=10^4-10^6$). 
We cautiously note that the averaging procedure will affect the detouring factor; $\langle B_y/B_z\rangle$ is different from $\langle B_y\rangle / \langle B_z\rangle$. When we compute the values in the brackets, we first horizontally average the data and take a time average 
for 50 orbital periods from the point of 350 orbits after the
beginning of the simulation. 

Taking the effect of detouring into consideration, 
we estimate the effective column density that cosmic rays have to go through to enter into the disk. 
The results are presented in the bottom panel of Figure~\ref{fig:By_Bz} for the disk around IM Lup \citep{zha21} as an example of the observed protoplanetary disks.
With conventional treatment, cosmic rays of attenuation length of $\sim100$ g cm$^{-2}$ \citep[][see \citet{pad18} for more updated values for the high-density region]{ume81}
can arrive at the midplane 
for $r\gtrsim$ 10 au unless the shielding effect by T-Tauriosphere is effective. 
On the other hand, our treatment indicates that cosmic rays cannot arrive at the 
disk midplane even at $\sim$ 100 au for IM Lup. 

In our simulations, the toroidal component of magnetic fields is generated at all heights
due to the ideal MHD approximation. 
In the very dense region with poor coupling with magnetic fields, i.e., near the midplane around a few au from the star, the field lines would not be stretched by the velocity shear. Then, cosmic rays do not have to detour much.
In such a region, however, cosmic rays are easily shielded, and our conclusions are unchanged.

%%%%%%%%%%%%%%%%%%%%%%%%%%%%%%%%
\section{Effects on the ionization rates}
               \label{sec:zeta}
%%%%%%%%%%%%%%%%%%%%%%%%%%%%%%%%
%%%%%%%
\begin{figure}
%\epsscale{0.6}
   \centering
   \plotone{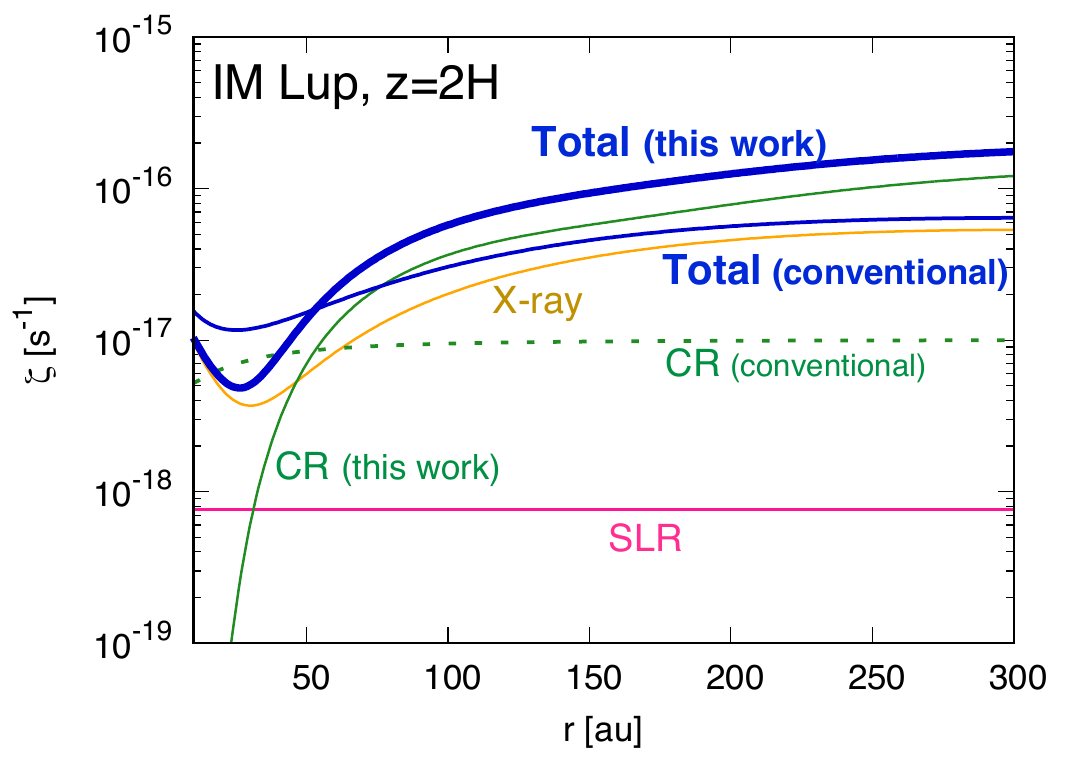}
   \plotone{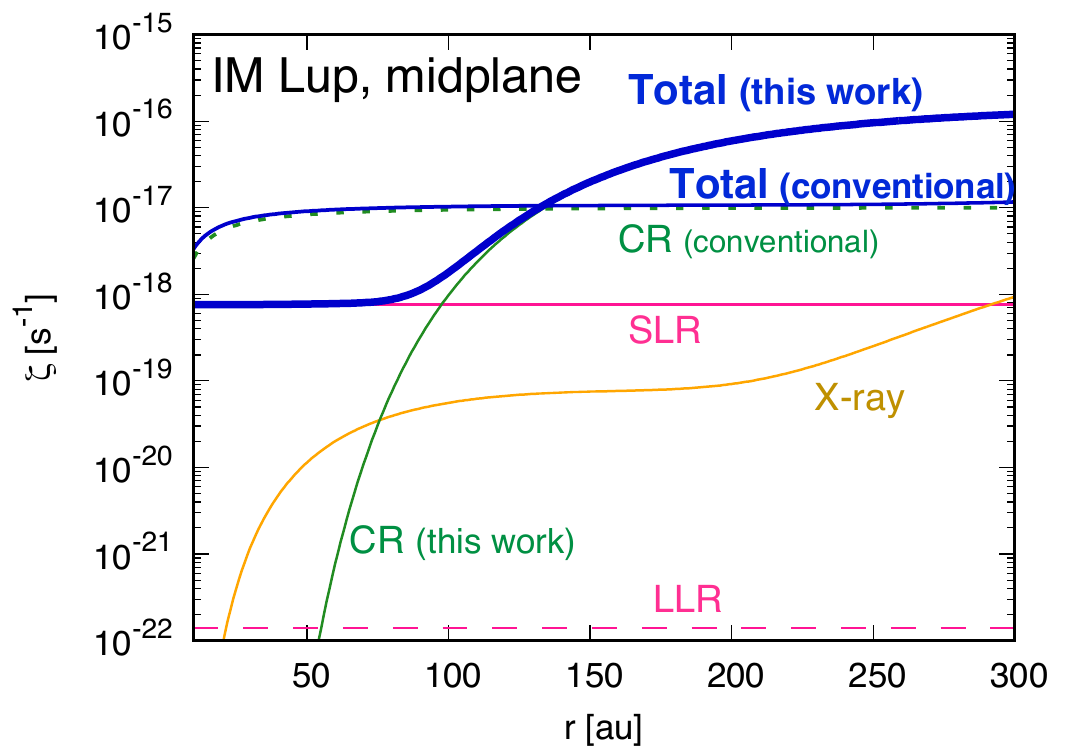}
   \plotone{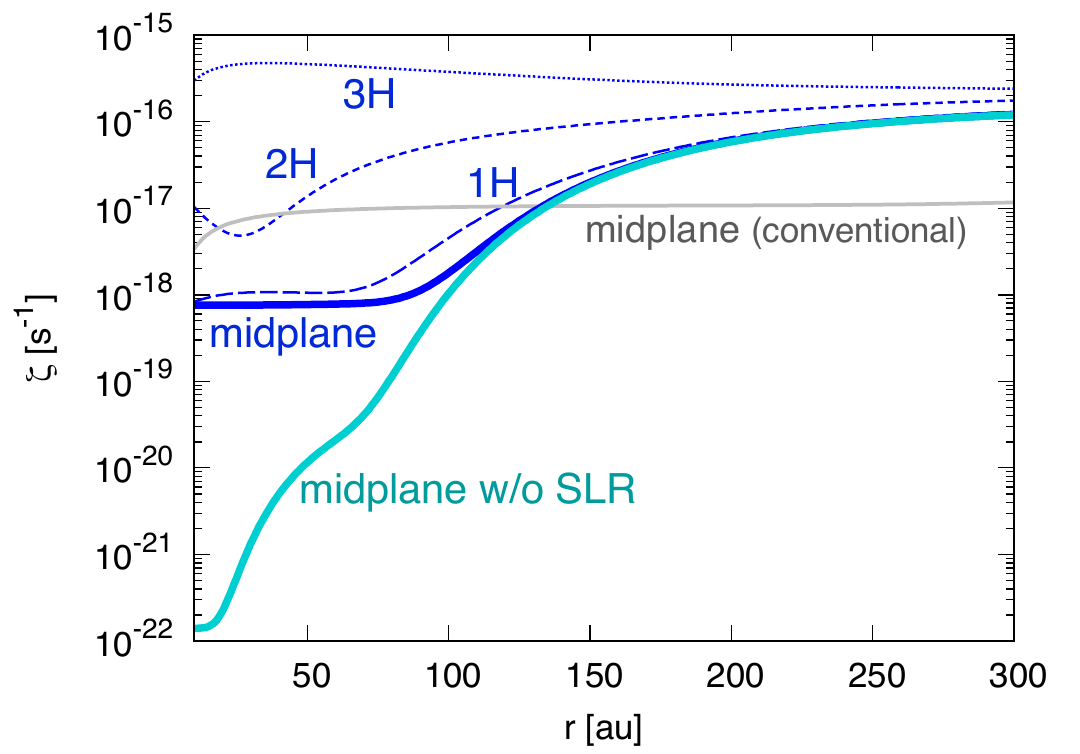}
   \caption{Top panel: the radial distribution of ionization rate at $2H$ 
       from the midplane based on the new estimate of the CR
       ionization rate (solid green) with $\zeta_{\rm ISM}=3\times10^{-16}$ s$^{-1}$ using the effective column density
       along the elongated path, together with the contribution of X-ray (orange) and short-lived radionuclides (SLR; pink) is displayed 
       with the thick blue curve. The conventional estimate of $\zeta_{\rm CR}$
       based on $\zeta_{\rm ISM}=10^{-17}$ s$^{-1}$ with $\Gamma_{\rm detour}=1$ (the dotted green line) makes up the thin blue curve of the 
       conventional ionization rate. 
       Middle panel: the same with the top
       panel, but for the midplane of the disk. The dashed pink line is the 
       ionization rate of the long-lived radionuclides (LLR).
       Bottom panel: the total ionization rate at $z=3H,\ 2H,\ 1H$, and midplane using the CR ionization rate of this work. The case with the absence of SLR and the estimate using the conventional CR ionization rate are plotted in light blue and gray, respectively. The x-axis in each plot starts from 10au.
           }
           \label{fig:zeta}
    \end{figure}
%%%%%%%
%%%%%%%
\begin{figure}
   \centering
   \plotone{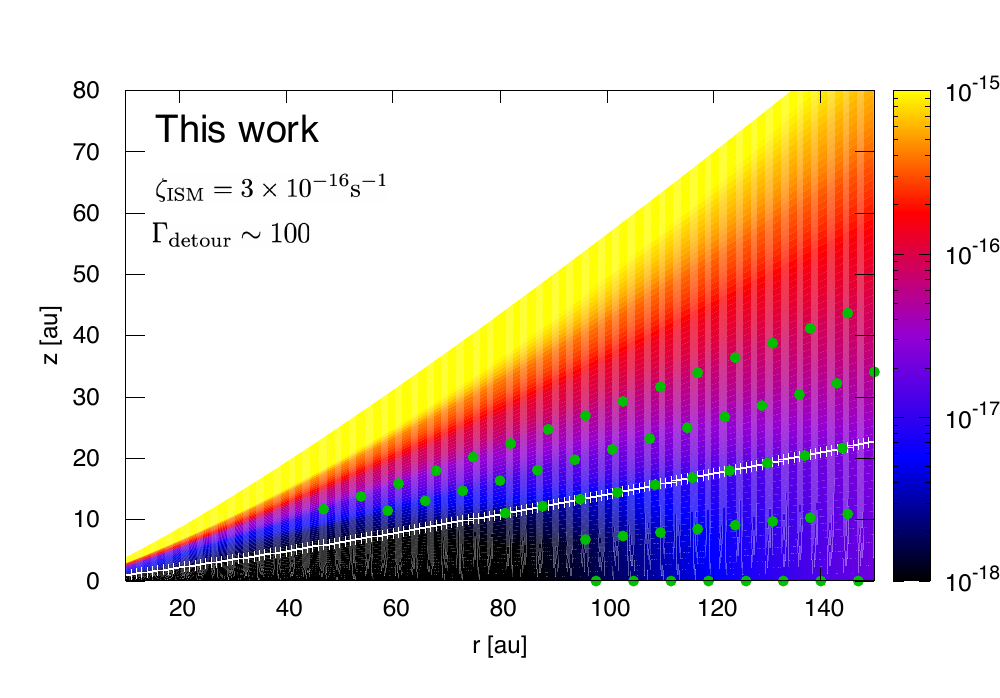}
   \plotone{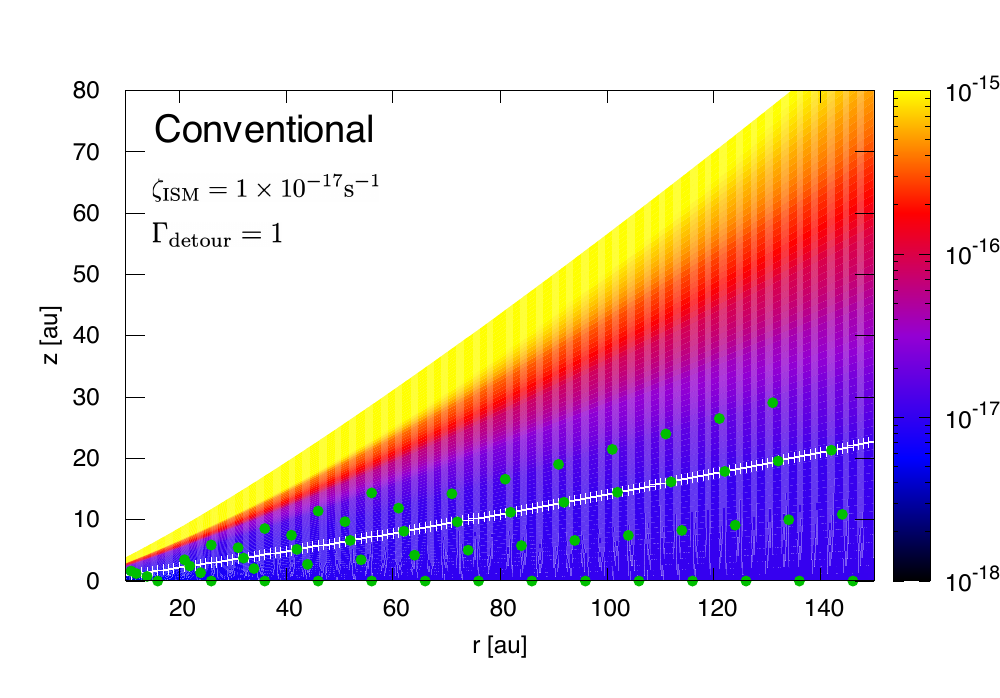}
   \caption{The distribution of ionization rate for our new model considering the propagation of cosmic rays along magnetic fields and that for the conventional model considering the vertical attenuation of cosmic rays in the disk around IM Lup. The white lines represent the scale height of the disk (which is $\sqrt{2}$ times larger than the definition of \citet{zha21}) and the green dots mark the regions where the cosmic rays are the dominant ionization source. 
      }
           \label{fig:zeta_dominant}
    \end{figure}
We show a significant enhancement of the effective column density 
in Figure~\ref{fig:By_Bz}. Due to the detouring of the cosmic rays along the sheared field lines, cosmic ray particles stay in the surface layer for a longer time before they enter into the disk.
The cosmic-ray ionization rate at the point of interest within the disk can be calculated as
\begin{eqnarray}
    \zeta_{\rm CR} = \frac{1}{2}\Gamma_{\rm dim}\zeta_{\rm ISM}
    \left\{ \exp({-\Gamma_{\rm detour} \chi_{\rm eff,u}/\chi_{\rm CR}})\right. \nonumber\\
    \left. + \exp({-\Gamma_{\rm detour} \chi_{\rm eff,l}/\chi_{\rm CR}}) \right\},\label{eq:zeta_cr}
\end{eqnarray}
where $\zeta_{\rm ISM}$ is the ionization rate of
cosmic rays in the interstellar medium
and $\chi_{\rm CR}=96$ g cm$^{-2}$
is the attenuation length of cosmic rays \citep{ume81}.
We define the vertical column density from the point to the upper/lower
surface of the disk as $\chi_{\rm eff,u}$ and $\chi_{\rm eff,l}$, respectively,
and $\Gamma_{\rm detour}$ is the ratio of the effective and vertical column density from the 
point to the closer surface, i.e. $\Gamma_{\rm detour}=1$ in conventional studies and $\Gamma_{\rm detour}\chi_{\rm eff}=\int \langle \rho B_y/B_z\rangle dz$ in our work.
Our results suggest $\Gamma_{\rm detour}\sim 100$ when the cosmic rays travel along the sheared magnetic fields.
 Intuitively, one would think that the CR density can be enhanced by the sheared magnetic field by a factor of $\Gamma_{\rm detour}$. However, the sheared magnetic field makes the mirroring effect effective, and thus, the CR density does not change (see Section \ref{sec:2}).

As an example, we estimate the ionization rate assuming the surface density and temperature distributions of the disk around IM lup \citep{zha21}. 
We consider 4 ionization sources: cosmic rays, X-rays, and short-lived (SLR) and long-lived radionuclides (LLR). The cosmic-ray ionization rate is given by Equation (\ref{eq:zeta_cr}). 
The typical value used in conventional studies is $\zeta_{\rm ISM}=10^{-17}$ s$^{-1}$ \citep{spi68, cum16}, and we adopt this for the conventional model. Observations of diffuse clouds in Galactic disk suggest higher values \citep{pad20}, however, those values are not consistent with lower ionization rate in the midplane of protoplanetary disks \citep{Aikawa21maps, sei21}. Here, we employ $\zeta_{\rm ISM}=3\times10^{-16}$ s$^{-1}$\citep{ind15} for our new estimates.
The value of $\Gamma_{\rm detour}$ is insensitive to the averaging method; the calculations of $\langle \rho\  B_y/B_z\rangle$ and $\langle \rho \rangle \langle B_y/B_z\rangle$ lead to similar results.  We use the X-ray ionization rate by \citet{bai11} that considers both the direct and scattered emissions based on \citet{ige99}. We use X-ray luminosity $L_X=10^{30}$ erg/s \citep{cle17} and plasma temperature $T_X=3$ keV. The ionization rate by SLR and LLR
are $\zeta_{\rm SLR}=7.6\times10^{-19}\rm~s^{-1}$ and $\zeta_{\rm LLR}=1.4\times10^{-22}\rm~s^{-1}$,
respectively \citep{ume09}.

The top panel of Figure~\ref{fig:zeta} shows the ionization rates as a function of $r$ at $z=2H$. We can see that
cosmic rays are attenuated above $z=2H$, and the X-ray ionization is dominant for $r \lesssim 50$ au. On the other hand, the conventional treatment leads to the opposite trend: the X-ray ionization is dominant at outer radii, while cosmic-ray ionization is important in the inner part of the disk.

The modification of the ionization rate in the midplane is summarized in the
middle panel of Figure~\ref{fig:zeta}. In the conventional model, cosmic rays are the dominant ionization source for all the radii.
On the other hand, the cosmic-ray ionization rate of our model at $r\lesssim 100$ au is much lower than that of the conventional model because of the significant reduction of the cosmic-ray flux. The ionization is dominated by SLR there.
The X-ray ionization is sub-dominant in all radii, because of the significant attenuation.

$\zeta_{\rm ISM}$ higher than the conventional model 
does not affect the total ionization rate in the upper layers of $z\gtrsim3H$, because the X-ray ionization rate is higher at the upper layer of the disk.
When the X-ray flux is much smaller, the 
ionization rate may be
dominated by the adopted $\zeta_{\rm ISM}$ even in the higher atmosphere of the inner disk.

The half-life of the most effective species among SLR, $^{26}$Al, is 
7 $\times 10^5$ years, and they might not necessarily still be present in 
the protoplanetary disks of interest (See Section 5 for the discussion on the production of $^{26}$Al). 
Thus, we calculate the ionization rate of 
the disk in the absence of the SLR (the bottom panel of Figure~\ref{fig:zeta}).
The ionization is actually led by LLR 
within $\sim$ 20 au in this case. 
Then, X-rays play an important role
for 20 au $\lesssim r \lesssim$ 70 au, and cosmic rays are dominant for $r\gtrsim70$ au.
\citet{ume09} pointed out 
that the ionization by radionuclides suddenly becomes inefficient when 
the mean size of the dust grains grows to about $>1$ cm because the radiation
is absorbed by the grains. In such a case, ionization would be governed by X-rays while cosmic rays are highly attenuated.

The 2D map of the ionization rate for our new model and that for the conventional estimate are shown in Figure~\ref{fig:zeta_dominant}. The ionization rate for our model is lower than that for the conventional model by about an order of magnitude for $z\lesssim 1H$ and $r\lesssim 100$ au due to the detouring of cosmic rays. The ionization rate in this region is dominated by SLR in our model, while the cosmic-ray ionization is dominant in the conventional model (see the distribution of green dots in the figure).
On the other hand, the ionization rate for our model is higher than that for the conventional model for $z\sim 2H$ or the midplane for $r\gtrsim120$ au, owing to 
the larger value of $\zeta_{\rm ISM}$.

%%%%%%%%%%%%%%%%%%%%%%%%%%%%%%%%
\section{Discussion}
               \label{sec:discussion}
%%%%%%%%%%%%%%%%%%%%%%%%%%%%%%%%

The suppression of the energy density of CRs with energies $<10$ GeV is observed in the solar system \citep{Pot13}. The situation in the protostellar system is more speculative. Some claim that winds from young stars should suppress the CRs of similar or even higher energies \citep{coh12,rod20b}. However, we consider that the T-Tauri winds are unlikely to affect the CR density in the protoplanetary disks. The mass outflow rate from T-Tauri stars is much lower than the winds from protoplanetary disks, and then, the T-Tauri winds cannot push out the disk winds. Thus, the disk winds fill the region above the disk and may modulate the CR density. We speculate that the magnetic field configuration is crucial for the suppression of CRs.  
If the disk winds are driven by the magnetic centrifugal force \citep{bla82,bai16}, the magnetic field lines anchored in the disk can be connected to the ISM, and the magnetic field lines  are not completely dominated by the toroidal component. Then, the cosmic rays are likely able to enter the disk by the free-streaming or diffusion parallel to the magneitc field lines. On the other hand, if the disk winds are driven by other mechanisms, such as turbulence heating \citep{suz10} or photoevaporation \citep{owe10,kim16ppd, kun20, kun21}, the magnetic field lines are unlikely connected to the ISM and likely dominated by the toroidal component. Then, the cosmic rays need to cross the magnetic field lines many times to reach the disk, which may suppress low-energy cosmic rays as in the solar winds.

Stellar energetic particles are regarded as a possible ionization source in protoplanetary disks. Depending on the settings of calculations,
 such as the particle energy and the treatment of propagation,
 they may be the dominant/important ionization source in the surface layer of the disk \citep{tur09, rab17, fra18} or even at the midplane of the disk \citep{rod20}.
 However, even for high energy particles of $E\sim10$ GeV, the Larmor radius is much smaller than the scale height, and they need to detour along the sheared magnetic fields as long as the diffusion perpendicular to the magnetic field lines are inefficient. Based on our scenario, stellar energetic particles may significantly contribute to the ionization rate at the disk surface, but they are unlikely to reach the midplane of the disk. Recent studies suggest $^{26}$Al can be produced by  stellar energetic particles \citep{gac20, ada21}. However, since the stellar energetic particles cannot reach the disk midplane, the newly synthesized $^{26}$Al is unlikely to affect the ionization rates in the range of our interest.

An ALMA large program, `Molecules with ALMA at Planet-forming Scales' (MAPS) discussed the ionization rates at the midplane of protoplanetary disks \citep{Aikawa21maps}.
Their results suggest $\zeta\sim10^{-18}\rm~s^{-1}$ and also show the variation among disks of a similar mass range. This might be caused by the different configurations of magnetic-field lines or the turbulent state of the disks. 
Another ALMA observation by \citet{sei21} reported a lower ionization rate in the inner disk and a higher value in the outer radii with the transition at $\sim$80--100 au in the disk around IM lup. Our model predicts a similar tendency as seen in Figure \ref{fig:zeta_dominant}. 

Some protoplanetary disks show lower values of the CO abundance compared to the canonical ISM value of $10^{-4}$ in the layer between the CO photodissociation layer in the surface and the CO freeze-out layer near the midplane \citep[e.g.,][]{fav13, mio17}. This may suggest a high ionization rate of $\zeta \geq10^{-17}$s$^{-1}$ and low temperature ($T\sim 20-30$\,K) in the disks because CO molecules are converted into less volatile molecules in such an environment \citep{fur14, bos18, sch18}. Our model with the large value of $\zeta_{\rm ISM}$ predicts higher ionization rates at the middle layer ($\sim 2 H$) in the outer disks ($\sim$100 au) compared to the canonical cosmic-ray ionization rate, while keeping the midplane ionization rate as low as values suggested by ALMA observations.

\begin{figure}
   \centering
   \plotone{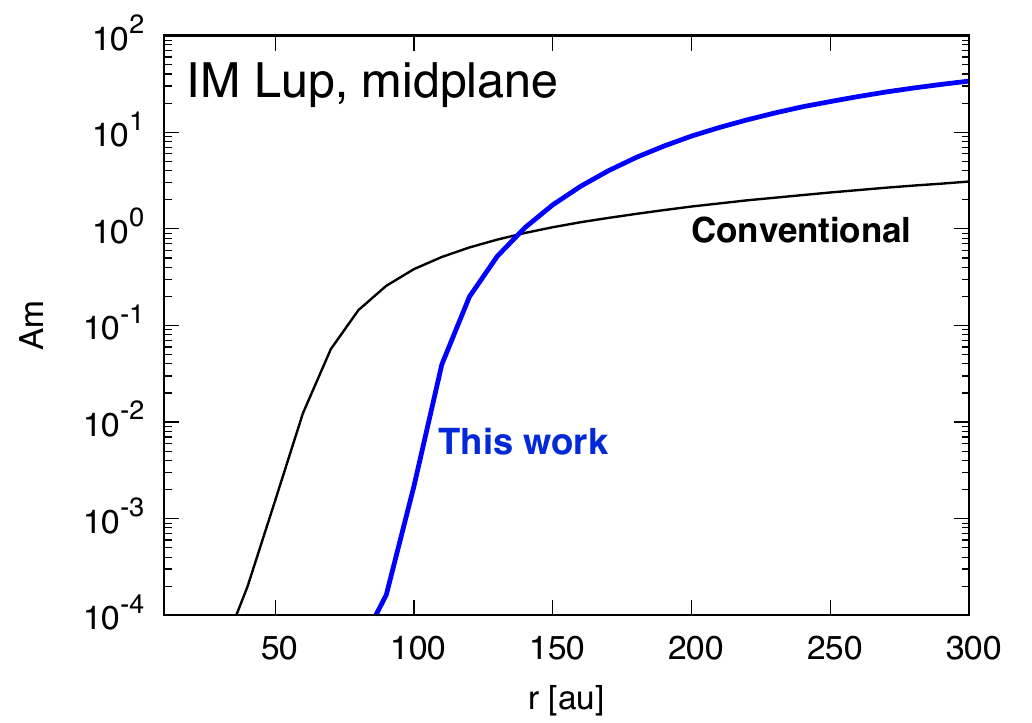}
   \caption{Radial distributions of Ambipolar Elsasser number at the midplane of the disk around IM Lup.
      }
           \label{fig:Am}
    \end{figure}

\citet{bai13b, bai13c, sim15b} showed that ambipolar diffusion weakens the MRI turbulence at the outer radii of disks. We calculate ambipolar Elsasser number Am $=v_A^2/(\eta_A \Omega)$, where $v_A$, $\eta_A$,  $\Omega$ are Alfven velocity, ambipolar diffusivity, and Keplerian frequency, respectively. We use the formulae of \citet{war07} for $\eta_A$ and compute the ionization rate with the simplified method of \citet{fuj11}. Figure~\ref{fig:Am} shows Am in the IM Lup disk. Our ionization model suggests Am $\lesssim1$, where we expect a purely laminar disk, within $\sim 150$ au. The tendency that the disk is laminar at $\sim$ 100 au is consistent with the recent ALMA observation suggesting very weak turbulence due to the small scale height of the dust disk in the disk around Oph 163131 \citep[e.g.,][]{vil22}. In the further outer disk,
where previous studies predict weak MRI turbulence due to ambipolar diffusion \citep{bai13b, bai13c, sim15b}, our model expects 
Am $\gtrsim$ 10. This may fully activate the MRI turbulence. Depending on the location, the Hall effect might be the dominant non-ideal MHD effect \citep{war12}. Non-ideal MHD simulations with our ionization model are required to further investigate the level of turbulence.

The existence of turbulence affects cosmic-ray transport. We have assumed that the magnetic fields have no fluctuations and cosmic-ray transport is in the free-streaming regime. If some mechanism, such as MRI, drives turbulence, fluctuations of magnetic fields can scatter the cosmic rays. Then, the cosmic-ray transport should be in the diffusion regime. Even in the diffusion regime, cosmic rays move along the magnetic field lines because they have an anisotropic diffusion coefficient with the parallel diffusion much faster than the perpendicular diffusion, as shown in particle simulations in MRI turbulence \citep{kim16cr,kim19,sun21}. Then, the particle trajectory can be expressed by the random walk along the field line. Thus, the path length to the disk midplane in the diffusion regime should be much longer than in the streaming regime. In the diffusion regime, cosmic rays likely lose their energy at a further upper region of the disk, and then, the ionization degree around the midplane of the disk would become low. This leads to suppression of turbulence, and cosmic rays would become able to enter the disk again.
In summary, we start the discussion from the laminar magnetic field (implying suppression of turbulence). Then, cosmic rays will enhance the ionization rate, and the disk will become the ideal MHD regime that drives turbulence. The turbulence would shield the cosmic rays, which leads to the non-ideal MHD regime in which turbulence is suppressed.
To consistently understand the evolution of protoplanetary disks, we need to solve the magnetohydrodynamics, cosmic-ray transport, and ionization rate simultaneously.

\section{Summary}\label{sec:sum}

We have investigated the effects of the magnetic-field configurations on cosmic-ray ionization rates in protoplanetary disks. We have first considered the cosmic-ray propagation from ISM to the disk, where magnetic field strength changes by several orders of magnitude (see Figure~\ref{fig:core_ppd}). The cosmic-ray density at the disk surface should decrease by a factor of $\Gamma_{\rm dim} \sim 1/2$ from the ISM value because of the balance between the magnetic-flux focusing and reflection by a magnetic mirror.

The magnetic fields are stretched to the azimuthal direction in  the disk due to the shearing motion, which affects the cosmic-ray distributions in the protoplanetary disks. Cosmic rays need to detour to reach the disk midplane due to the shearing magnetic field, 
and the vertical penetration depth should decrease. 
We have performed a set of local MHD simulations to obtain the detouring path length and showed that the cosmic-ray density 
at midplane for $r\lesssim 100$ au can be significantly suppressed even when we employ $\zeta_{\rm ISM}=3\times10^{-16}$ s$^{-1}$. Our results suggest a totally different ionization rate distribution from that of the conventional model. This may change the chemical and dynamical evolution of the protoplanetary disks, which is crucial to understanding the star and planet formation processes.

\begin{acknowledgments}
We acknowledge Satoshi Okuzumi, Yusuke Tsukamoto, Jiro Shimoda, Yuri Aikawa, Kenji Furuya, Yoshihide Yamato, and Akimasa Kataoka for useful discussions. We are grateful to Soonyoung Roh and Marco Padovani for fruitful discussions at the beginning of this project. We also thank Kedron Silsbee and Alexei Ivlev for useful comments. This work is partly supported by JSPS KAKENHI Grant Number 19H05077, 22K14086 (Y.I.F), 19J00198, 21H04487, and 22K14028 (S.S.K.). Y.I.F was supported in-part by the Program for the Development of Next-generation Leading Scientists with Global Insight (L-INSIGHT), sponsored by the Ministry if Education, Culture, Sports, Science and Technology (MXT), Japan. S.S.K. acknowledge the support by the Tohoku Initiative for Fostering Global Researchers for Interdisciplinary Sciences (TI-FRIS) of MEXT's Strategic Professional Development Program for Young Researchers. Numerical computations were in part carried out on Cray XC50 at Center for Computational Astrophysics, National Astronomical Observatory of Japan.

\end{acknowledgments}

\appendix
\section{Minimum Mass Solar Nebula Model}
\begin{figure}
%\epsscale{0.6}
   \centering
   \plotone{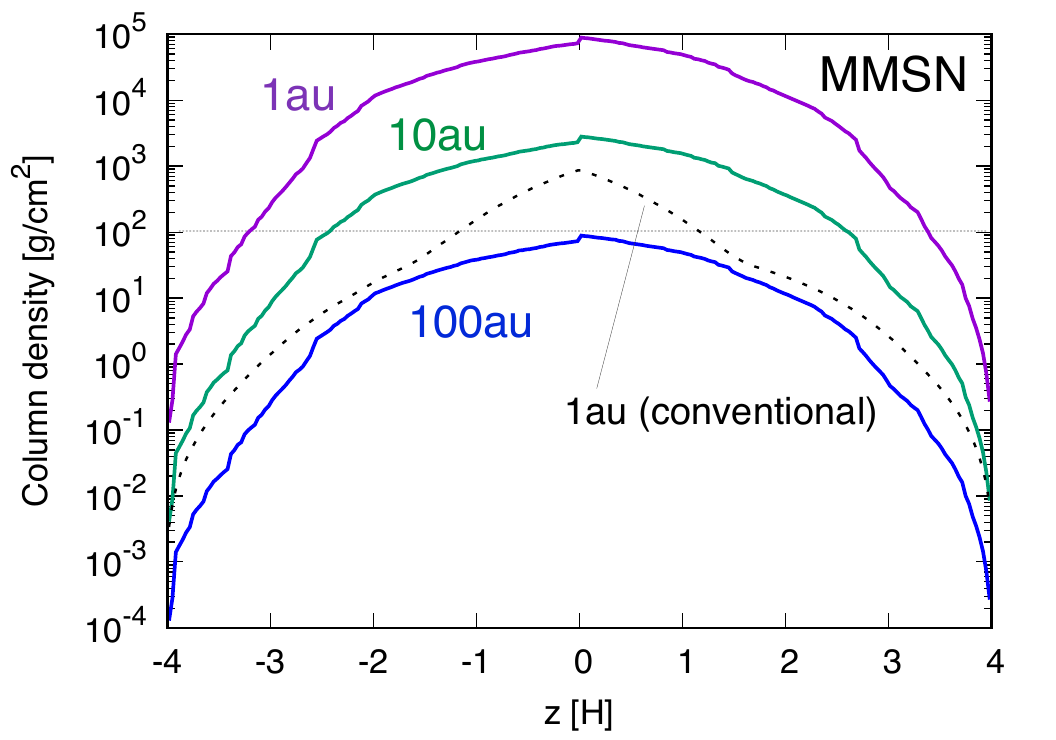}
   \caption{
       The effective column densities that cosmic rays have to pass through
       from the closer surface to reach each height of the MMSN at 1, 10, and 100au.
       The dashed black line is the vertically measured column density that is used in conventional studies for 1au. The thin horizontal line shows the attenuation length of cosmic rays.
   }
               \label{fig:MMSN}
    \end{figure}

For the comparison with theoretical studies adopting the Minimum Mass Solar Nebula (MMSN) model \citep{hay81}, we provide the effective column density for cosmic rays to arrive at various heights from the midplane of MMSN. Conventionally, cosmic rays are thought to easily reach the midplane at $\sim$ 10\,au. As one can see in Fig.\ref{fig:MMSN}, however, they can barely penetrate into the midplane at 100\,au in our new model.
%%%%%%%%
%%%%%%%

\bibliography{manuscript}{}
\bibliographystyle{aasjournal}

\end{document}